# Upper Critical Field and Kondo Effects in Fe(Te $_{0.9}$Se$_{0.1}$) Thin Films by Pulsed Field Measurements

Myron B. Salamon [1,3,*], Nicholas Cornell [2], Marcelo Jaime [3], Fedor F. Balakirev [3], Anvar Zakhidov [1], Jijie Huang [4], and Haiyan Wang [4]

[1]Department of Physics, The University of Texas at Dallas, Richardson, TX 75080 USA
[2]Naval Surface Warfare Center, Corona Div., Norco, CA 92860
[3]National High Magnetic Field Laboratory, Los Alamos National Laboratory, Los Alamos, NM 87545
[4]Department of Materials Science, Texas A&M University, College Station,TX 77843-3128
[*]salamon@utdallas.edu

**ABSTRACT**
The transition temperatures of epitaxial films of Fe(Te $_{0.9}$Se$_{0.1}$) are remarkably insensitive to applied magnetic field, leading to predictions of upper critical fields $B_{c2}(T = 0)$ in excess of 100 T. Using pulsed magnetic fields, we find $B_{c2}(0)$ to be on the order of 45 T, similar to values in bulk material and still in excess of the paramagnetic limit. The same films show strong magnetoresistance in fields above $B_{c2}(T)$, consistent with the observed Kondo minimum seen above $T_c$. Fits to the temperature dependence in the context of the WHH model, using the experimental value of the Maki parameter, require an effective spin-orbit relaxation parameter of order unity. We suggest that Kondo localization plays a similar role to spin-orbit pair breaking in making WHH fits to the data.

Magnetism and superconductivity are intimately connected. In the context of S-wave order parameters, magnetism was considered to be detrimental to superconductivity, so much so that one of the famous Matthias Rules[1] cautioned that magnetic elements were to be avoided when seeking new materials. The situation changed with the advent of heavy-Fermion superconductors[2] and, likely, the cuprates[3] where magnetism provides the pairing mechanism. With the discovery of iron-based superconductors,[4] the Mathias Rule may well have been an impediment to their earlier discovery, rather than a helpful guide. The simplest of the Fe-based superconductors is the series Fe(Te,Se), which crystalize in simple iron-chalcogenide layers. In this paper, we focus on thin epitaxial films of Fe(Te $_{0.9}$Se$_{0.1}$); the epitaxial strain suppresses the tetragonal-to-monoclinic transition and antiferromagnetism found at this concentration in bulk FeTe.[5] A previous report[6] has demonstrated the quality of these films, grown by pulsed-laser deposition on SrTiO$_3$. One of the intriguing aspects of that prior work is the insensitivity of the transition temperature to applied field, giving an initial slope $dB_{c2}=dT$ as large as -16 T/K measured at the midpoint of the transition, which is near $T^{mid}_c \approx$ 11 K in zero field. The usual WHH expression[7] $B_{c2}(0) = -0.69Tc (dB_{c2}/dT)_{Tc}$ predicts a value $B_{c2}(0)$ = 130 T. In this work, we report the actual upper critical field using pulsed magnetic fields at the National High Magnetic Field Laboratory. Further, we note that these films typically show a resistance minimum above the superconducting transition consistent with a Kondo effect. An STM study by Yin et al.[8] found that superconducting Fe(Te,Se) films exhibit zero-bias anomalies at the always-present interstitial Fe sites. Although Yin et al. considered, but could not conclude, that these anomalies are Kondo bound states, we assume that interstitial Fe atoms act as Kondo scatterers. We find that samples with resistance minima also exhibit strong negative magnetoresistance above $B_{c2}$ which can also be analyzed in terms of the field dependence of the Kondo effect.[9, 10]

Samples were grown at Texas A&M University by a method described previously.[6] The $SrTiO_3$ crystalline substrates were cleaved to appropriate size (approximately 3 x 5 mm$^2$) for 4-terminal resistivity measurements in a Quantum Design Physical Properties Measurement system, for R(H,T) measurements to 2K and in dc fields up to 7 T. The same samples were mounted on a cryostatic insert at the National High Magnetic Field Laboratory at Los Alamos National Laboratory for resistivity studies to 0.7 K in pulsed magnetic fields to 60 T. Copper leads for both experiments were connected to the 100 nm films using silver paint. Care was taken to twist current and voltage leads to minimize inductive pickup from the pulsed field. However, evidence for Joule heating of the films was evident from slightly different R(H,T) curves on up-field and down-field sections of the magnet pulse. Samples in the pulsed-field experiment were immersed in liquid helium below 4 K; some problems with sample heating were encountered when the sample is in exchange gas at temperatures between 4 K and the transition temperature $T_c \approx$ 12 K, a challenging temperature range due to small sample heat capacity and poor thermal conductivity of the exchange gas. A preliminary report of this work was contained in the PhD dissertation of N. Cornell.[11]

Figure 1 shows the pulsed field data taken on Sample 018 with the field normal to the sample film; other samples gave identical results. The data in the normal state show significant negative magnetoresistance which follows a uniform curve as a function of temperature and field. We will return to the treatment of this effect below. The field dependence fits smoothly with the low field data taken in dc fields, shown in Figure 2.

Above Tc, the samples generally exhibit a minimum R(T); followed by a logarithmic increase in resistance with decreasing temperature, as shown in Figure 3. The red line in the inset corresponds to a fit with a Kondo temperature $T_K$ = 60 K, and an increase of 38 $\Omega$/decade. Kondo impurities in superconductors have been studied extensively, most notably by Maple and coworkers[12] and theoretically by Muller-Hartmann and Zittartz[13] and by Matsuura, et al.[14] A key point of the theory, which we discuss below, is that the behavior of a superconductor with magnetic impurities differs from standard BCS[15] results when $T_K >> T_c$. We are concerned here with the behavior of the Kondo contribution to the low-temperature resistivity at fields such that $m_B H \sim k_B T_K$. An empirical treatment of the Kondo resistance in fields, proposed by Felsch and Winzler,[9] gives the following expression,

$$R(H,T) = R_{bkg} + r_K \left[1 - \frac{\ln(T/T_K)}{\left(\ln^2(T/T_K) + \pi^2 S(S+1)\right)^{1/2}}\right]\left[1 - B_S^2\left(\frac{g\mu_B S H}{k_B(T+T_K)}\right)\right] \qquad (1)$$

where the impurity spin is S and $B_S(x)$ is the Brillouin function for spin S. We focus on the field-dependent term, since $T/T_K$ is less than 0.15 over the temperature range of interest. We test this behavior by replotting some of the data shown in Figure 1, rescaling the field in each curve by *(1+T/T$_K$)*, as shown in Figure 4. The solid line is a fit to the field-dependent term in (1) with S = 1; the temperature dependent term in (1) varies between -0.37 and -0.5 over the temperature range. We ignore the temperature dependence in our fit, so that our value of RK is approximately 1.4$r_K$.

The conclusion here is that excess Fe present in epitaxial Fe(Te$_{0.9}$Se$_{0.1}$) films produces Kondo resistivity in the normal state which is partially suppressed at 60 T, which corresponds to 82 K for S = 1.

Figure 5 shows the field-temperature points at which the resistance is 50% of the maximum resistance in the curves in Figure 1. The circles are for fields normal to the film; squares, in the plane of the film, and triangles, dc data with B normal. We find that $B^{mid}_{c2}(0)$ = 43 T for both orientations of the field; the initial slope $(dB^{mid}_{c2}/dT)_{Tc}$ = -16 T/K, measured in dc fields, is also independent of field direction. This differs from single-crystal data on Fe(Te$_{0.6}$Se$_{0.4}$) in which $B_{c2}(0)$ is independent of field direction, but the initial slope is larger for fields in the ab-plane.[16] The $B_{c2}(T)$ curve in the c-direction is more linear than our data here, although the zero-temperature value is the same. Fang et al.[16] suggest that ripples in the Fermi surface cause it to be more three dimensional and make $B_{c2}(0)$ more isotropic.

The usual approach to discussing the upper critical field relies on the WHH theory.[7] In the absence of Pauli limiting, the usual estimate for orbital pair-breaking gives $B^{orb}_{c2}(0)$ = 132 T slope -16 T/K of the mid point of the transition from Figure 2. This value greatly exceeds the paramagnetic pair-breaking limit[17] $B_p[T] = \sqrt{2}\Delta/\mu_B$ = 12.2$\Delta$[meV]; the gap in Fe(Te$_{0.85}$Se$_{0.15}$) was measured by Kato, et al.[18] and found to be $\Delta$ = 2.3 meV, so that $B_p$ = 28 T. A further refinement, suggested by Schlossman and Carbotte.[19] modifies the Clogston result to read $B_p = \sqrt{2}\Delta\eta_\lambda/\mu_B$ that allows for the electron-phonon mass-enhancement coefficient $\lambda$ and possible Stoner enhancement of the susceptibility, from Fe impurities, perhaps. As seen in Figure 5, the actual upper critical field (midpoint of the transition) is $B^{mid}_{c2}$ = 43 T. The usual treatment, due to Maki, is to introduce the parameter $\alpha$ which interpolates between the orbitally limited and paramagnetically limited fields according to

$$B^{mid}_{c2}(0) = \frac{B^{orb}_{c2}(0)}{\sqrt{1+\alpha^2}} \tag{2}$$

This leads to the value $\alpha_{exp}$ =3. However, the Maki parameter $\alpha$ is actually defined as $\alpha_{Maki} = \sqrt{2}B^{orb}_{c2}(0)/B_p\eta_\lambda$ =6.7/$\eta_\lambda$ ; we do not expect $\eta_\lambda$ , which takes into account the competing effects of electron-phonon mass enhancement and Stoner increase in susceptibility, to differ strongly from unity. In either case, these values of a, when included in the WHH expression, lead, absent spin-orbit effects, to re-entrant behavior.[7] The value of the Maki parameter above which the WHH expression for $B_{c2}(T)$ becomes double-valued is given by

$$\alpha_c = \frac{1+1.589\lambda_{SO}/\lambda^c}{1-\lambda_{SO}/\lambda^c} \tag{3}$$

where $\lambda^c$ = 0.539 is determined numerically in Ref.(8). For $\alpha_{exp}$ to give rise to a single-valued curve, we must have $\lambda_{SO} \geq 0.2$; for $\alpha_{Maki}$, we must have $\lambda_{SO}$ = 0.37. When the spin-orbit parameter is larger than $\lambda^c$, the $B^{mid}_{c2}(T)$ curve is always single-valued.[7]

In Figure 5, we show two curves for $\alpha_{exp}$ and $\lambda_{SO}$ = 0 (dashed) and for $\alpha_{Maki}$, with $\eta_\lambda$ = 1 and $\lambda_{SO}$ = 1.2 treated as an adjustable parameter. Clearly, the dashed curve is double valued, as expected, for $\alpha \geq 1$ in the absence of spin-orbit scattering. We note that $B^{orb}_{c2}$ and, consequently, the value of $\alpha_{Maki}$, increases with increasing slope $(dB^{mid}_{c2}/dT)_{Tc}$. Samples of the same composition exhibit a wide range of initial slope, but very similar values for $B^{mid}_{c2}(0)$.[20-22] Within the context of the WHH model, the curves can only match the initial slope and $T$ = 0 value if there is a variation in $\lambda_{SO}$ among samples. As this is unlikely, we must consider alternative approaches.

As noted above, the Kondo effect significantly modifies the impact of magnetic impurities on superconductivity. When the Kondo temperature $T_K$ is significantly below the pure system superconducting transition temperature $T_{c0}$, the conventional Abrikosov-Gorkov[15] pair-breaking model holds, and the transition temperature decreases to zero with impurity concentration. However, for $T_K \gg T_{c0}$, impurities are Kondo-screened and pair-breaking gives way to a repulsive interaction that reduces the superconducting interaction. In Matsuura et al.,[14] $B_{c2}(0)$ depends on the impurity concentration, as does the initial slope. The numerical results in Ref.(15) do not include the possibility of paramagnetic depairing, and are valid only when $B_{c2}(0) \ll B_K = k_B T_K/\mu_B$; neither condition being met in Fe(Te$_{0.9}$Se$_{0.1}$). Although quantitatively inapplicable to the present data, the model demonstrates that the initial slope of the $B_{c2}(T)$ line can increase with increasing impurity content and consequently so does the estimated value of $B^{orb}_{c2}(0)$ and the correspondingly larger value of the Maki parameter. Indeed, preliminary data on another sample, labeled S021, show both a smaller initial slope $(dB^{mid}_{c2}/dT)_{Tc} \approx 12T/K$ and a smaller magnetoresistance effect (see Eq. 1) as a percentage of the base resistance than do either samples S018 or LC247, as reported here.

Binary iron chalcogenides are not simple single-band superconductors,[24] so fitting them into one-band, BCS theory is obviously overly simplified. Two band fits have been moderately successful for single crystals of FeSe[23] but involve a large number of adjustable parameters without providing a better representation of the critical field. We choose, for the purposes of this work, to regard the spin-orbit parameter as a sample-dependent parameter, possibly correlated with the excess Fe concentration, which could be monitored via the magnitude of the Kondo resistance $R_K$. This would require measurements on samples with a variety of initial values of $(dB^{mid}_{c2}/dT)_{Tc}$ and therefore high-field, low temperature measurements on such samples.

**Acknowledgments**
The NHMFL Pulsed Field Facility is supported by the National Science Foundation, the U.S. Department of Energy, and the State of Florida through NSF Cooperative Grant No. DMR-1157490. This work was also supported by the US Air Force Office of Scientific Research contract FA9550-09-1-0384 on "Strengthening superconductivity in macro-arrays of nanoclusters and nanostructures".

## Author contributions

MBS performed high-field measurements, analyzed data and prepared figures, wrote manuscript. NC performed high-field and low field measurements, carried out preliminary analysis, work included as part of his dissertation. MJ Performed high-field measurements, made key editorial suggestions for improvements in figures and text. FB Participated and supported high-field measurements, assisted with cryogenics in pulsed fields. AZ Supervised low field measurements and analysis, edited manuscript. JH Prepared thin film samples, carried out preliminary low-field analysis. HW Supervised film synthesis, provided editorial suggestions. All authors reviewed the manuscript in advance of submission.

## Additional Information

The author(s) declare no competing financial interests

## Figures

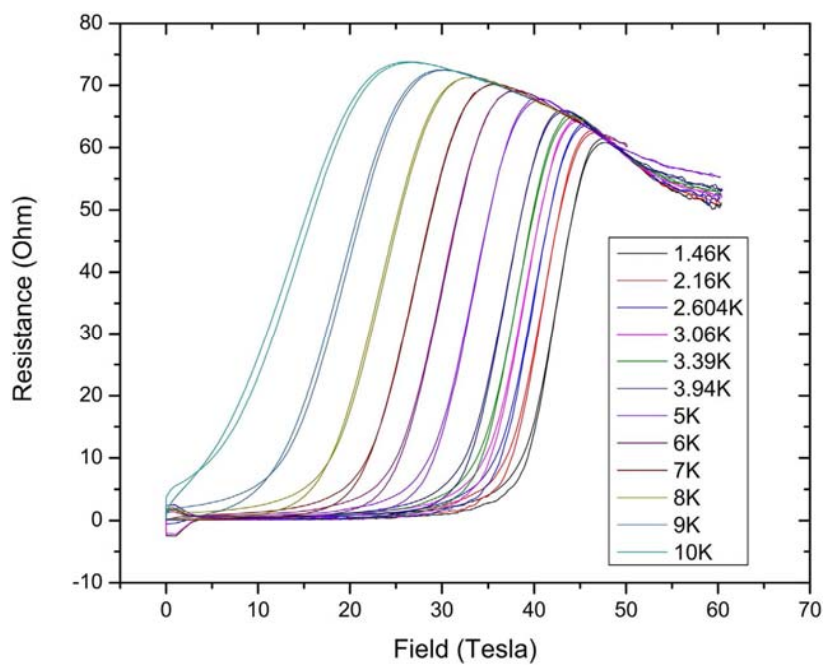

**Figure 1. Pulsed field data on Sample 018.** Note the large negative magnetoresistance in the normal state, which follows a general trend.

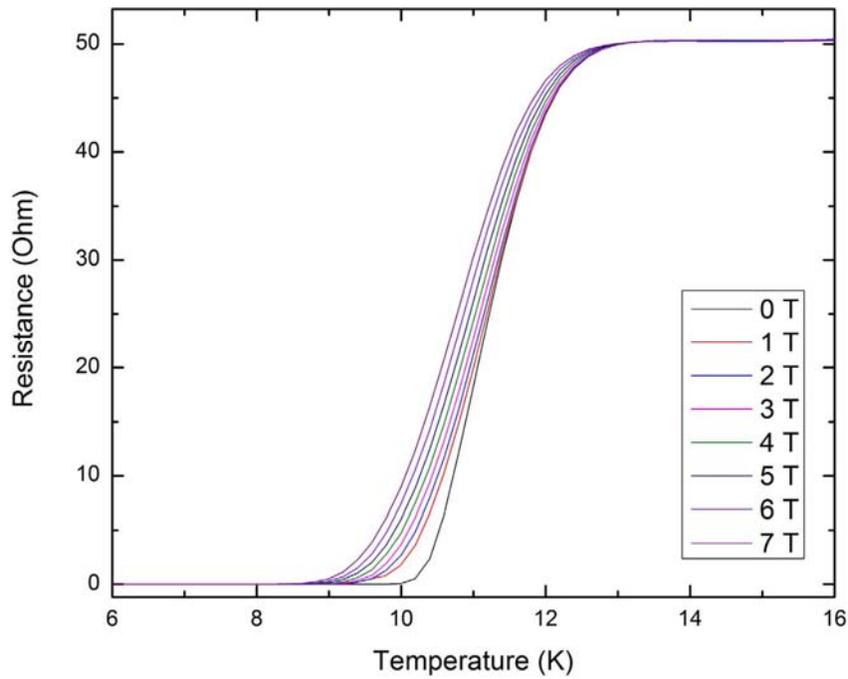

**Figure 2. Resistive transition in dc fields for sample S018.** Note the relative insensitivity of the transition to applied field, with the midpoint moving only about 0.5 K in 7 T.

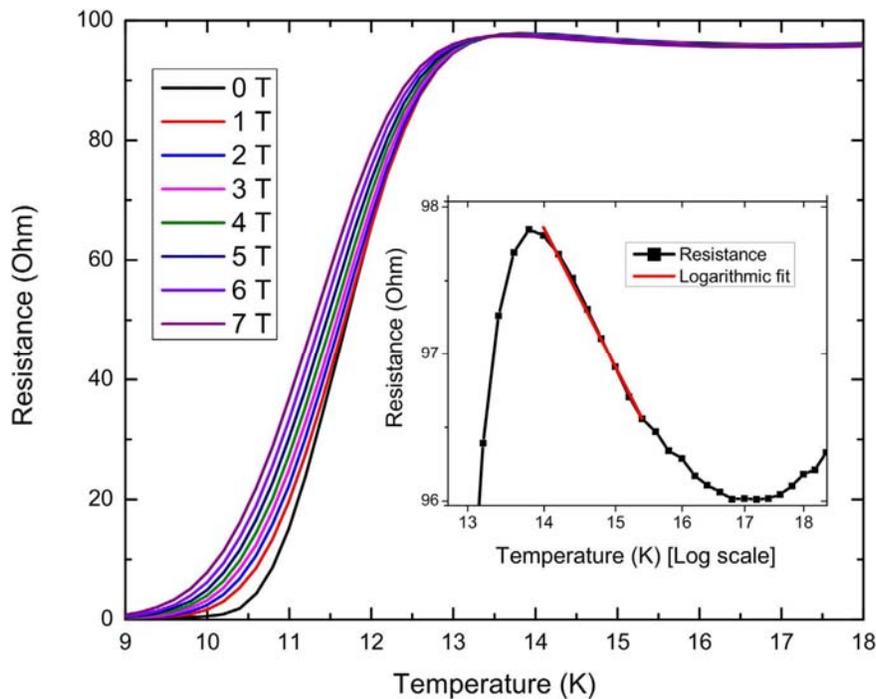

**Figure 3. Field dependent transition for sample LC247.** The inset shows the resistance in the normal state in zero field, plotted on a logarithmic temperature scale. The straight line is a logarithmic fit with a Kondo temperature of 60 K and a slope of 38 Ω/decade.

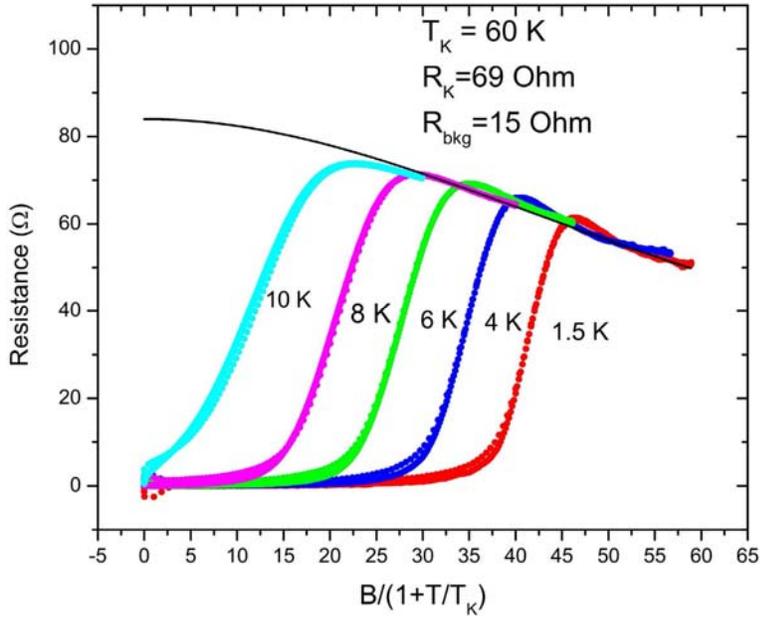

**Figure 4. Data from Fig. 1, scaled by reduced temperature for each curve.** The solid line is a fit to Eq. 1, with the fitting parameters indicated on the Figure.

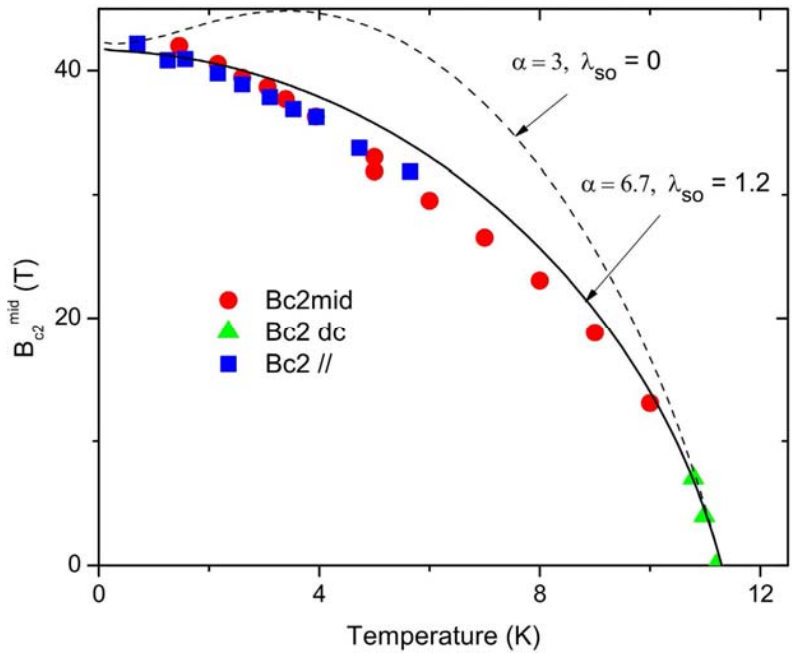

**Figure 5. Midpoint upper critical fields. Circles: pulsed fields normal to film; triangles: dc fields normal to film; squares: fields in the plane of the film.** The dashed curve is a WHH fit with $\lambda_{SO}$ = 0, and the solid curve is with $\alpha_{Maki}$ and $\lambda_{SO}$ as an adjustable parameter.